\documentclass{article}

\usepackage{microtype}
\usepackage{graphicx}
\usepackage{subfigure}
\usepackage{booktabs} 

\usepackage{hyperref}
\usepackage{xspace}

\usepackage{algorithm}
\usepackage{algorithmic}

\usepackage[accepted]{icml2025}

\usepackage{amsmath}
\usepackage{amssymb}
\usepackage{mathtools}
\usepackage{amsthm}
\usepackage{tikz}
\usetikzlibrary{calc, positioning, shapes.geometric, arrows.meta}
\usepackage[dvipsnames]{xcolor}
\definecolor{diagramblue}{HTML}{a6caec}
\definecolor{diagramgreen}{HTML}{b4e5a2}

\usepackage[capitalize,noabbrev]{cleveref}

\theoremstyle{plain}

\theoremstyle{definition}

\theoremstyle{remark}

\usepackage[textsize=tiny]{todonotes}

\newcommand{\OurSys}{\textsc{DistributedANN}\xspace}
\newcommand{\DiskANN}{\textsc{DiskANN}\xspace}

\icmltitlerunning{\OurSys}

\begin{document}

\twocolumn[
\icmltitle{\OurSys: Efficient Scaling of a Single \DiskANN Graph Across Thousands of Computers}



\icmlsetsymbol{equal}{*}
\icmlsetsymbol{atmsft}{\textdagger}

\begin{icmlauthorlist}
\icmlauthor{Philip Adams}{microsoft}
\icmlauthor{Menghao Li}{stca}
\icmlauthor{Shi Zhang}{microsoft}
\icmlauthor{Li Tan}{microsoft}
\icmlauthor{Qi Chen}{msra,equal}
\icmlauthor{Mingqin Li}{shopify,atmsft,equal}
\icmlauthor{Zengzhong Li}{microsoft,equal}
\icmlauthor{Knut Risvik}{trondheim,equal}
\icmlauthor{Harsha Vardhan Simhadri}{microsoft,equal} 

\end{icmlauthorlist}

\icmlaffiliation{microsoft}{Microsoft, Redmond, United States}
\icmlaffiliation{stca}{Microsoft, Bejing, China}
\icmlaffiliation{trondheim}{Microsoft, Trondheim, Norway}
\icmlaffiliation{shopify}{Shopify, Bellevue, United States}
\icmlaffiliation{msra}{Microsoft Research Asia, Vancouver, Canada}

\icmlcorrespondingauthor{Philip Adams}{philipadams@microsoft.com}

\icmlkeywords{Information Retrieval, Approximate Nearest Neighbor Search, Distributed Systems, Search Engine Scalability}

\vskip 0.3in
]



\renewcommand{\icmlEqualContribution}{\textsuperscript{*}Listed alphabetically by surname\quad\textsuperscript{\textdagger}The work was done at Microsoft.}

\printAffiliationsAndNotice{\icmlEqualContribution} 

\begin{abstract}
We present \OurSys, a distributed vector search service that makes
it possible to search over a single 50 billion vector graph index
spread across over a thousand machines that offers $26$ms median query latency 
and processes over 100,000 queries per second.
This is $6 \times$ more efficient than existing partitioning and
routing strategies that route the vector query to
a subset of partitions in a scale out vector search system.
\OurSys is built using two well-understood components:
a distributed key-value store and an in-memory ANN index.
\OurSys has replaced conventional scale-out architectures
for serving the Bing search engine, and we share our
experience from making this transition.
\end{abstract}

\section{Introduction}
 Approximate Nearest Neighbor (ANN) search  is a common retrieval technique in web search, multimedia search, and new scenarios like Retrieval-Augmented Generation \cite{lewis2020retrieval}. Given a set of data vectors $X$, and a query vector $q$, the goal of an ANN search system is to quickly find as many of the $k$ vectors in $X$ that are nearest to $q$ as possible. ANN search is a classic problem and researchers have developed many techniques, including algorithmic approaches \cite{HNSW2018IEEE,andoni2008near,Suhas2019NEURIPS,Chen2018Github,babenko2014inverted}, compression techniques \cite{jegou2010product,Ge2014TPAMI,gao2024rabitq}, and specialized indexes targeting GPUs, flash storage, or external memory \cite{johnson2019billion,ootomo2024cagra,ChenW21NIPS,Jang2023ATC}. The performance of these techniques is closely tracked by benchmarking efforts like ANN-Benchmarks \cite{annbenchmark} or Big-ANN-Benchmarks \cite{bigann2023}.

 However, the majority of research effort has been focused on datasets small enough to fit in a single machine's memory or disk. For larger datasets, like searching over hundreds of billions of web documents, the standard approach is to split the corpus into smaller partitions\footnote{For online systems, it is often necessary to choose a partition size much smaller than the total space available on a machine, in order to be able to bring a new replica online quickly enough should a host fail.}. An independent index is built for each partition, and each partition is hosted on a different machine so that the entire corpus can be searched in parallel. The main downside of this approach is efficiency: inside a single ANN index, query cost
  scales with $\log \lvert X \rvert$ (as empirically measured), while across many partitions (assuming a fixed partition size) the cost will scale with the number of partitions.
 In other words, a system with P partitions has a search complexity 
 of $P \log \frac{\lvert X \rvert}{P}$, which is much worse than the complexity
 of one index over the entire dataset $X$.
 Even techniques aimed at improving this tradeoff, like using clustering to assign vectors to partitions \cite{ChenW21NIPS}, face scalability challenges and complex tradeoffs.

 \begin{figure*}[htb!]
          \begin{tikzpicture}[node distance=1.5cm and 2cm]

  \node at (0,3.5) {Clustered Partitioning};
  \node at (9,3.5) {\OurSys};

  \draw[dashed, thick] (4.5,-1) -- (4.5,4);
  
  \node[fill, fill=orange!20, text width=8cm, align=center, inner ysep=2mm] (select) at (0,2.5) {Select best \textbf{N}  partitions};
  
  \node[fill, fill=diagramgreen, text width=2cm, align=center] (cache1) at (-3,1) {Cache};
  \node[fill, fill=diagramgreen, text width=2cm, align=center] (cache2) at (-.5,1) {Cache};
  \node[fill, fill=diagramgreen, text width=2cm, align=center] (cache3) at (3,1) {Cache};
  
  \node[fill, fill=diagramblue, text width=2cm, align=center, inner ysep=1.5mm] (graph1) at (-3,0.05) {Graph partition \\ (\textbf{M} SSD IO)};
  \node[fill, fill=diagramblue, text width=2cm, align=center, inner ysep=1.5mm] (graph2) at (-.5,0.05) {Graph partition \\ (\textbf{M} SSD IO)};
  \node[fill, fill=diagramblue, text width=2cm, align=center, inner ysep=1.5mm] (graph3) at (3,0.05) {Graph partition \\ (\textbf{M} SSD IO)};
  
  \node at (1.25,0.25) {$\ldots$};
  
 
  \draw[-{Latex}] (select) -- (cache1);
   \draw[-{Latex}] (cache1) -- (select);
  \draw[-{Latex}] (select) -- (cache2);
  \draw[-{Latex}] (cache2) -- (select);
  \draw[-{Latex}] (select) -- (cache3);
  \draw[-{Latex}] (cache3) -- (select);
  
  
  \node[fill, fill=orange!20, text width=8cm, align=center, inner ysep=2mm] (orch) at (9,2.5) {Orchestration Service};
  
  \node[fill, fill=diagramgreen, text width=2.5cm, align=center, inner ysep=4mm] (head) at (6.25,0.25) {Head \\ Index \\ Search};
  \node[fill, fill=diagramblue, text width=4.5cm, align=center, inner ysep=4mm] (global) at (10.75,.25) {Global \\ Graph nodes \\ (\textbf{BW} SSD IO)};
  
  \node[align=center] (hgraph) at (10.75,1.75) {$\mathbf{H}$ graph hop\\ batches};
  
  \draw[-{Latex}] (orch) -- (head);
    \draw[-{Latex}] (head) -- (orch);
  \draw[-{Latex}] (orch) -- (global);
  \draw[-{Latex}] (global) -- (orch);

\end{tikzpicture}
     \caption{High-level architecture comparison between a conventional system using clustered partitioning and \OurSys.}
     \label{fig:architecture}
 \end{figure*}
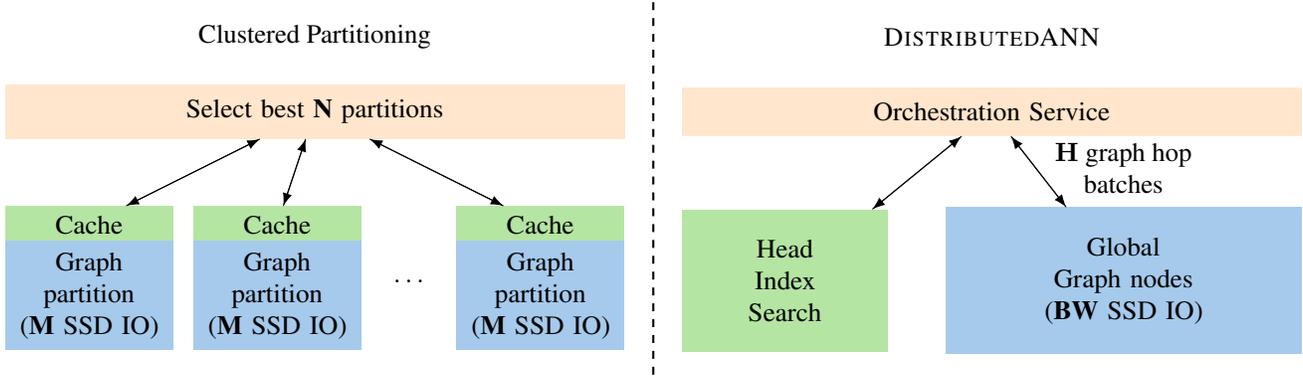

  In this paper, we argue that state-of-the-art performance on very-large-scale datasets can be achieved through a single large logical index stored in a distributed key-value store. \OurSys begins with the abstraction of the key-value store as a large shared disk, and then makes  modifications to the data and compute placement choices of \DiskANN indices in order to make it practical to serve them in this setting. We explore the tradeoffs compared to a traditional approach, and present experimental results demonstrating the scalability of this approach on a fifty-billion vector subset of a web search dataset. \OurSys has already been deployed in Microsoft Bing to support search over hundreds of billions of vectors. Compared to the previous production system, it delivered over \textbf{6$\times$ headroom in query throughput} with the same machine footprint, and \textbf{7.8 and 4.5 percentage point improvements in recall@5 and recall@200} respectively.

\section{\OurSys}
Existing ANN indices like \DiskANN are optimized for tiered storage by reducing the number of round-trips to SSD. 
Therefore, \DiskANN is well suited for adaptation to a distributed serving environment, since the main drawback of accessing a networked disk is added access latency. However, directly serving a multi-hundred-terabyte \DiskANN graph from network-attached storage encounters new bottlenecks:
\begin{itemize}
    \item The memory-resident portions of the index are too large to fit in a single machine, and accessing them over the network would incur significant overhead.
    \item As the diameter of the graph grows, more hops are needed to traverse it, increasing latency. 
    \item Transmitting the full graph node data over the network requires too much serial compute per query and consumes excessive network bandwidth.
\end{itemize}  
\vspace{-7pt}

In the rest of this section, we first review the data layout of \DiskANN, and then discuss how we adapt the index to overcome these scaling bottlenecks. 

\subsection{\DiskANN Index Layout}
A \DiskANN index has 3 components: 
\begin{enumerate}
    \item An array of compressed (PQ or OPQ\footnote{Product Quantization and Optimized Product Quantization, described in detail in \cite{jegou2010product,Ge2014TPAMI}.}) representations of all vectors in the index, stored in memory.
    \item An array of graph nodes, one per vector, stored on SSD. Each node stores the full precision vector and a list of IDs representing out-neighbors.
    \item An in-memory cache of frequently-accessed graph nodes.
\end{enumerate}

Searching this index begins at a fixed point, and scores neighbor candidates of each visited node using their in-memory compressed representations to determine which nodes to visit in the next iteration of beam search. Depending on the index size and the recall required, tens to hundreds of nodes may be read (depending on $I$, the limit on the total amount of IO), and for each node many candidate neighbors (depending on $R$, the graph degree) will be considered.  
\subsection{Index Layout Modifications}
\paragraph{Compressed Vectors Duplicated into Graph Nodes.}
Our first modification is based on the observation that, for a sufficiently large index, the array of compressed vectors will not be able to fit in a single machine. One option would be to store these compressed vectors in a memory-based key-value store. However, due to the large number of candidates that must be considered ($I\times R$ may be tens of thousands of lookups per search for typical parameters) and the latency implications of doing an extra network hop per beam search iteration, we instead decide to duplicate the compressed representation of each vector into all the graph nodes it is a neighbor of. This introduces a significant space amplification\footnote{\label{footnote:spaceamp}For parameters of $R=100$, $d=384$, $d_{\operatorname{OPQ}} =64$ and using 8-byte IDs instead of 4-byte IDs to allow an index of more than 4 billion vectors, this is approximately a $10$x amplification.} of 
\begin{equation}
\label{eqn:spaceamp}
    \frac{(1+R)\operatorname{sizeof}(\operatorname{id})+d + Rd_{\operatorname{OPQ}}}{R\operatorname{sizeof}(\operatorname{id})+d},
\end{equation}
 but it reduces the number of read operations per search to $I$. This approach mirrors that of \cite{tatsuno2024aisaq,pan2023lm}.

\paragraph{In-Memory Head index.}
We also observe that while a node cache in the key-value store still has the benefit of reducing IO operations, it does not provide the same latency benefit as in single-machine \DiskANN because even cached nodes will still incur network hop latency when read. This is a major issue, since the shortest path to the furthest node from the starting point of the graph has at least $\log_R \lvert X \rvert$ edges. In conventional \DiskANN, the first few beam search iterations almost always hit the node cache, reducing the number of round trips to disk and thus the latency. To achieve a similar effect in \OurSys, we introduce a dense cache of the top layers of the graph. We first conduct a breadth-first traversal to collect $C$ vectors from the top layers of the \DiskANN graph. We then build a conventional sharded in-memory ANN index over these vectors. We call this smaller index the \textit{head index}. At search time, we first search in the in-memory head index, and then use the results as the starting points for beam search of the \DiskANN graph.

\subsection{Near-data Computation}

Our next modification is also motivated by latency. While log scaling within a single index is favorable compared to linear scaling across indices, it will still require $\log100 \approx 6$ times as much computation to achieve similar quality searching a 50-billion vector index as a 500-million vector one. If we treat the key-value store purely as a virtual disk and do this work sequentially in a single machine like in \DiskANN, there will be a corresponding increase in latency. We instead introduce a near-data node scoring service running on each key-value host, described in Algorithm~\ref{alg:nodescore}. 

\begin{algorithm}[hb]
   \caption{Node Scoring Service}
   \label{alg:nodescore}
\begin{algorithmic}
   \STATE {\bfseries Input:} Node keys $\{k_i\}$, threshold score $t$, candidate limit $l$, full-dimension query $q$, SDC encoded query $q_{\mathrm{SDC}}$ 
   \STATE {\bfseries Static Data:} OPQ distance table
   \STATE {\bfseries Output:} Sorted result IDs and distances $R$, sorted candidate IDs and distances $C$
   \STATE Initialize $R\leftarrow\varnothing, C\leftarrow\varnothing$
   \STATE Batch read the node entries $n_i$ for all $k_i$
   \FORALL{$n_i$}
   \STATE Compute $d(q,v)$ for full-dimension vector $v\in n_i$ and insert $v$ into $R$
   \FORALL{OPQ candidate $p\in e_i$}
   \IF{$d_\mathrm{OPQ}(q_{\mathrm{SDC}}, p) < t$}
   \STATE Insert $p$ into $C$
   \ENDIF
   \ENDFOR
   \ENDFOR
   \STATE Sort $R$ and partial-sort $C$ up to $l$
   \STATE Truncate $C$ to $l$
\end{algorithmic}
\end{algorithm}

This change has the benefit of parallelizing the computation, while also allowing resources to be consumed in small uniform chunks (each key read will incur a similar amount of scoring work\footnote{We observe that due to the high ratio of shards to $\operatorname{BW}$, the typical batch size of this service is $1$, resulting in very predictable resource usage.}) that work well with existing resource management/load balancing systems\footnote{Multiple scenarios search the web index, and each has a different ideal resource vs. quality tradeoff. By consuming resources in roughly equal chunks, the existing key-value store load balancer can transparently accommodate different search parameters. In a partitioned index, different search parameters required separate benchmarking to set appropriate load factors for each scenario.}. Additionally, since we only transmit scores over the network instead of full nodes, we achieve a bandwidth savings\footnote{Using the same parameters as in Footnote~\ref{footnote:spaceamp}, this is approximately a 6x saving. We increase the savings further by pruning any neighbors that are worse than the current worst member of the candidate heap before returning to the orchestration service.} of 
\begin{equation}
\label{eqn:bandwidthsaving}
    \frac{(1+R)\left(\operatorname{sizeof}(id)+\operatorname{sizeof}(score)\right)+d+d_{\mathrm{OPQ}}}{(1+R)\operatorname{sizeof}(\operatorname{id})+d + Rd_{\operatorname{OPQ}}}
\end{equation}
compared to a naive virtual disk approach. 
\subsection{Orchestration Service}

The final component of \OurSys is an orchestration service that maintains lists of the best seen results and candidate vectors. This service will first issue a search in the head index, and then issue $H$ rounds of calls to the node scoring service before returning a final set of results to the caller, described formally in Algorithm~\ref{alg:orch}. 
\begin{algorithm}[ht]
   \caption{Orchestration Service}
   \label{alg:orch}
\begin{algorithmic}
   \STATE {\bfseries Input:} Full dimension query vector $q$. Beam width $\mathrm{BW}$. Beam iterations (hops) $H$. Result count $k$. Head index result count $k_\mathrm{head}$. Candidate size $L\geq\operatorname{max}(\mathrm{BW}, k)$.
   \STATE {\bfseries Static Data:} OPQ distance table, OPQ codebooks
   \STATE {\bfseries Output:} Sorted result IDs and distances $R$
   \STATE Initialize result heap $H_R$ of size $k$, candidate heap $H_C$ of size $L$. 
   \STATE Encode OPQ query $q_\mathrm{SDC}$ using the codebooks.
   \STATE Search for $k_\mathrm{head}$ results in the head index, and insert into $H_C$
   \FOR {$i=1$ {\bfseries to} $H$}
   \STATE Let $t = \operatorname{peekworst}(H_C)$
   \STATE Take best $\mathrm{BW}$ candidates from $H_C$ as keys $K$.
   \STATE Let $\{R_i\}, \{C_i\} = \operatorname{NodeScoring}(K, t, L, q, q_{\mathrm{SDC}})$.
   \STATE Partially merge-sorted-lists of $\{R_i\}$ upto $k$ and $\{C_i\}$ upto $L$, then insert into respective heaps.
   \ENDFOR
   \STATE Sort $H_R$ into $R$
\end{algorithmic}
\end{algorithm}
Because this service has a small amount of persistent state, it can be hosted on many machines with low overhead, ensuring that the load is evenly distributed. This service is also able to use hedged requests \cite{dean2013tail}, track replica health across requests, and allow partial failures of batches of node reads in order to reduce the tail latency normally associated with a high-fanout system like \OurSys. 

\section{Constructing a Large Graph Index}
\label{sec:unifiedgraph}
\begin{figure}[ht!]
    \centering
    \includegraphics[width=0.9\linewidth]{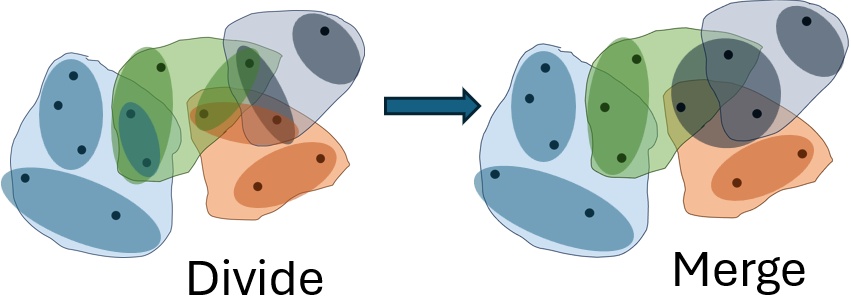}
    \caption{A visual depiction of the unified graph construction process. Partitions are represented by colored regions, and neighborhoods by dark shading. When points exist in multiple partitions, their neighborhoods will be merged by taking the union of neighbor lists, yielding a unified graph.}
    \label{fig:graphstitch}
\end{figure}

\DiskANN graphs are typically built incrementally by searching with the vector to be inserted as a query, tracking all the visited nodes during the search for use as potential graph neighbors, and then pruning to at most $R$ actual neighbors. While it is possible to insert vectors into an \OurSys graph by this procedure, it would require significantly more computation than building an equivalently sized partitioned graph, because the partitioned approach only needs to search in one smaller partition for each insertion. To reduce the graph construction cost, we employ a graph stitching approach similar to the one described in \cite{Suhas2019NEURIPS}. We first build an index with clustered partitioning, with vectors in the closure of multiple clusters inserted into all of them as described in \cite{ChenW21NIPS}. Because these vectors occur in multiple partition's graphs, we are able to stitch together a unified graph by taking the union of their neighbors from all the partitions they are present in, as shown in Figure \ref{fig:graphstitch}. In order to ensure that the entire graph is reachable, we build the head index from the union of the top layers of each partition's graph, rather than the top layers of the stitched-together graph. The quality of a graph built by this stitching process is lower than one built entirely incrementally, but is sufficient to get good results and is much faster to build.

\section{Evaluation}
The Bing web index is composed of multiple independent slices to allow portions of the index to be updated atomically using less capacity than would be required to update the entire index atomically. Each slice consists of roughly 50 billion 384-dimensional \texttt{int8} vectors. We compare the performance characteristics of \OurSys and a traditional clustered partitioning approach (with roughly 200 million vectors per partition) on one slice of the index. Because of the graph stitching approach described in Section~\ref{sec:unifiedgraph}, we are able to ingest identical indexes for both approaches, though for \OurSys we only ingest the first 72 neighbors in each node to reduce storage consumption. All experiments are conducted in a production environment which has a mix of host SKUs and multiple workloads sharing the resources of each host. A typical host in this environment has 256 to 768 \texttt{GiB} of memory, 5 to 10 \texttt{TiB} of SSD storage, roughly 200 IOPS per \texttt{GiB} of SSD storage, 32 to 64 physical cores, and 40 \texttt{Gbps} of network bandwidth. We ensure that the SKU mix hosting each system is roughly equivalent. The index parameters are chosen so that each system has a similar footprint (by bounding resource) at 15k QPS. The conventional index is bound by IO while \OurSys is bound by SSD space and can continue scaling to over 100k QPS in the same footprint. 
Performance data was collected on a set of sampled web search queries, presented in Table~\ref{tab:perfcompare}. The parameters for \OurSys are $H=5,\operatorname{BW}=128,R=72,k=L=200,k_\mathrm{head}=200$, with a head index size of $2.5$ billion vectors. The index built with clustered partitioning selects the top $40$ out of $203$ partitions, and searches in each with parameters $I=120,\operatorname{BW}=6,R=106,k=L=120$. 

\begin{table*}[ht]
    \centering
    \caption{Performance and accuracy comparison between \OurSys and conventional approach with 3 replicas}
    \label{tab:perfcompare}
    \begin{tabular}{l|cc}
        \toprule
        Metric & \OurSys & Clustered Partitioning \\
        \midrule
        Recall@5 (\%) & \textbf{90.8} &  83.0 \\
        Recall@200 (\%) & \textbf{71.9} & 67.4 \\
        Latency@50-ile (ms) & 26 &  \textbf{16} \\ 
        Latency@99-ile (ms) & 35 &  \textbf{22} \\
        SSD Space (\texttt{TiB}) & 780 & \textbf{270} \\
        Memory (\texttt{TiB}) & 42 & \textbf{18} \\
        IO per query & \textbf{640} & 4800 \\
        Network Bandwidth per query (\texttt{MiB}) & 1.4 & \textbf{0.3} \\
        Throughput (QPS) & $\mathbf{>}$\textbf{100k} & $\sim$15k \\
        \bottomrule
    \end{tabular}
\end{table*}
\subsection{Scaling}
\begin{figure}
    \centering
    \includegraphics[width=0.95\linewidth]{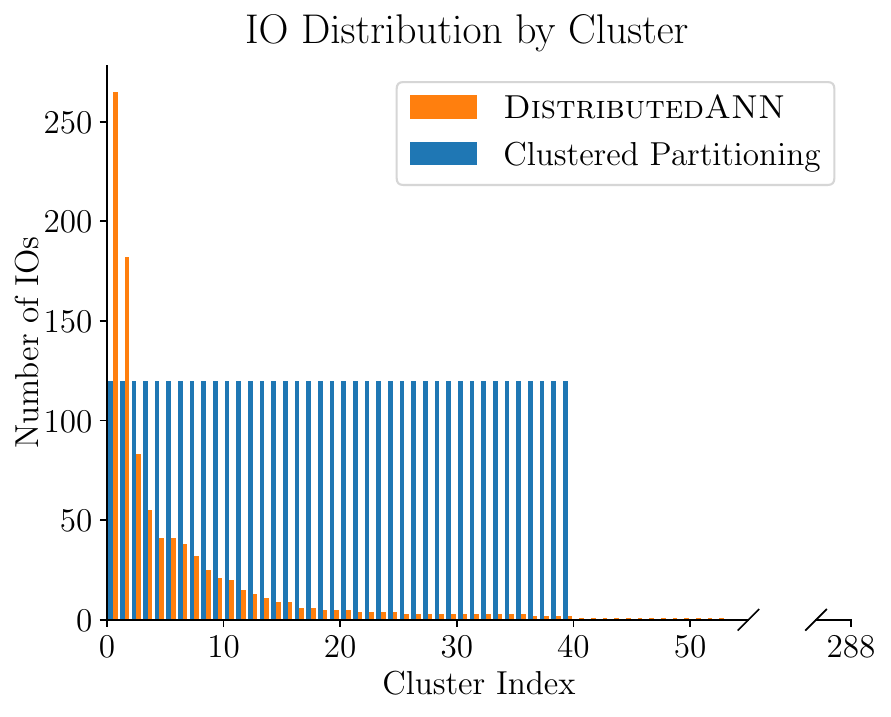}
    \caption{IO in each cluster for a single query, as served by \OurSys and a conventional clustered partitioning index. \OurSys is able to implement a much more flexible traversal strategy, improving efficiency.}
    \label{fig:io-comparison}
\end{figure}
\begin{figure}
    \centering
    \includegraphics[width=0.95\linewidth]{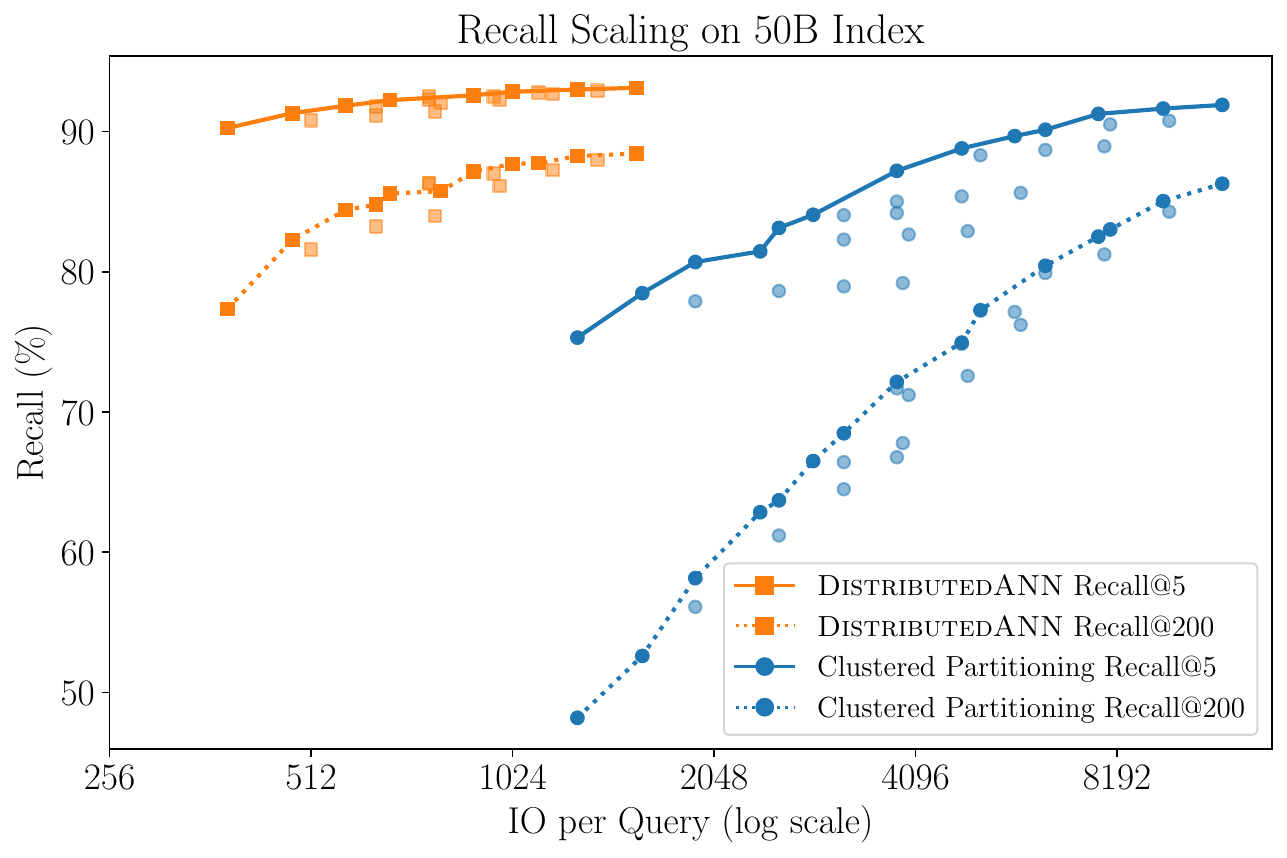}
    \caption{Optimal Recall/IO frontier for indexing 50 billion vectors. Grid search of parameters, \OurSys: $H$ from $4$ to $8$, $\operatorname{BW} = 32i$ for $i$ from $3$ to $6$. Clustered Partitioning: selected clusters $N=\left\{20,25,30,40,50,60\right\}$, IO per cluster $M=32i$ for $i$ from $2$ to $6$. }
    \label{fig:io-scaling}
\end{figure}

We observe that while \OurSys does consume significantly more storage space and has higher latency than a conventional system, it uses significantly less IO and is able to achieve higher throughput and recall on the same graph. Because \OurSys has unified graph, it can more efficiently allocate its IO budget, both by searching more deeply in relevant clusters and by touching more total clusters\footnote{Note that due to the graph stitching approach described in Section~\ref{sec:unifiedgraph}, a single read in \OurSys may touch multiple clusters.}. In comparison, a traditional clustered partitioning approach allocates a fixed amount of IO to each chosen partition (as shown in Figure~\ref{fig:io-comparison}). This allows \OurSys to achieve consistently higher recall across a range of IO budgets, shown in Figure~\ref{fig:io-scaling}.

\OurSys also enjoys more flexible and efficient throughput scaling than a conventional system, because the underlying key-value store can be sharded across more hosts to increase the available IOPS and CPU. In a conventional architecture, increasing the number of partitions would require more partitions to be searched for each query, increasing the IO per query. So, the main way to improve throughput is by increasing the number of replicas, which requires resources to be allocated in much coarser-grained chunks and consumes additional SSD and memory in addition to the extra IOPS and CPU.

\label{subsec:headindexbottleneck}
One significant bottleneck we identified in the current implementation of \OurSys is the head index, which at 3 replicas becomes CPU-bound before the node scoring service. Our solution in the current system is to increase the number of head index replicas. Due to the relatively small size of the head index and the bounding resource of the overall system (SSD space), this does not increase the number of machines required to serve the index.

\subsection{Reliability}
One potential concern for a distributed architecture is its resilience to issues in system components, like network partition or host failure. Although the mechanics of operating a high-availability distributed key-value store are well-understood in industry \cite{decandia2007dynamo}, \OurSys should still be resilient to partial failures in the node-storage layer. To test this resilience, we modified the node scoring service to accept a configurable failure rate parameter. We then examined the impact on recall@5 and recall@200 of  different failure rates, shown in Table~\ref{tab:availabilitydeg}.

The service experiences a graceful degradation in recall roughly proportional to the failure rate. This not only gives confidence in the reliability of the system, but also the performance stability, as we can safely timeout node scoring requests experiencing tail latencies without a significant adverse effect on recall. 

We contrast this with our experience operating a conventional partitioned ANN service. In such a service, availability is more difficult to maintain because of the higher partition sizes requiring more time to bring a new replica online. When a partition becomes unavailable, a large chunk of the dataset is missed, causing a dramatic drop in search quality. Additionally, when vectors are partitioned by clustering and only a subset of partitions are used for each search, load becomes imbalanced, as described in Subsection~\ref{subsec:clsimbalance}. The most popular clusters receive the most traffic and so are most likely to experience performance degradation, meaning that common queries will have the largest drop in search quality. In practice, independent partition scaling and heavy over-provisioning are required to achieve good availability, leading to lower resource utilization.

\begin{table}[h!]
    \centering
    \caption{Recall with degraded node scoring service availability}
    \label{tab:availabilitydeg}
    \begin{tabular}{c|cc}
        \toprule
        Availability (\%) & Recall@5 (\%) & Recall@200 (\%)\\
        \midrule
        100 & 90.8 &  71.9 \\
        99 &  89.7 & 70.1 \\
        98 &  88.8 & 69.4 \\
        97 &  87.5 & 68.8 \\
        96 &  87.0 & 67.8 \\
        \bottomrule
    \end{tabular}
\end{table}

\subsection{Comparison with GPU-based systems}
Recent work \cite{johnson2019billion,zhao2020song,ootomo2024cagra,khan2024bang} has sought to exploit the massive parallelism of GPUs to greatly increase the search throughput of a single machine. These systems have meaningly shifted the tradeoff between vertical and horizontal scaling of ANN search, and offer extremely competitive price-to-performance for high-throughput, billion-scale indices. However, these systems still have limitations on the size of index that can fit in a single machine, and eventually encounter the same linear scaling properties and operational complexities as other partitioned indices on very large datasets. For our scenario, the scalability benefits of \OurSys outweigh the advantages of GPU-based indices.  
\subsection{Comparison with Advanced Partitioning Schemes}
\label{subsec:clsimbalance}
Recent work such as \cite{Dong2019LearningSP,gotte2024arxiv} improves the performance of partitioned graph indices through schemes that reduce the number of cross-partition edges in an $k$-NN graph over the dataset. Further work is needed to compare the empirical performance of these approaches and \OurSys. In particular, these approaches may be preferable in very latency-constrained scenarios where the multi-hop network overhead of \OurSys is unacceptable. However, we feel that \OurSys has some notable operational advantages:
\begin{itemize}
    \item \OurSys can use one algorithm to build the entire index, rather than separate approaches for coarse-grained partitioning and fine-grained index building. Because of the incremental insertion approach of \DiskANN, build and search use the same code and can be optimized simultaneously.
    \item Semantic partitioning schemes are difficult to load-balance. Since queries will only access a subset of partitions, efficient serving requires independently scaling each partition with traffic, which is often difficult due to the different timescales between query pattern shifts and replica migration. By contrast, the underlying key-value store of \OurSys is randomly sharded and so receives a predictable traffic distribution.
\end{itemize}
\section{Conclusion}
\OurSys achieves sublinear scaling of ANN search on very large datasets, in exchange for reasonable increases in latency and disk space overhead. It also reduces operational complexity by reusing existing distributed key-value store infrastructure. We feel these tradeoffs are favorable and now use \OurSys in Microsoft Bing to enable search volume growth and quality improvements on the web index. 
\subsection{Future Directions}
We believe significant improvements in latency and space overhead are possible. Some opportunities include:
\begin{itemize}
    \item \OurSys uses traditional kernel-based TCP networking for remote service calls. Kernel-bypass networking would likely reduce the tail latency of these network calls significantly. More ambitiously, the relatively simple node scoring service logic could be implemented in a computational storage device attached to the orchestration host.
    \item As we note in Subsection \ref{subsec:headindexbottleneck}, the head index quickly becomes compute bounded in our tests. It may be cost-effective to serve this index from GPU rather than increasing the number of replicas to serve more traffic.
    \item Because traditional \DiskANN does not incur the space amplification that \OurSys does from replicating compressed vectors into each node, techniques for reducing the average number of graph edges per vector have not been explored in depth. A significant reduction in space overhead is likely possible by placing multiple nearby full-dimension vectors into a single graph node, but further experimentation is needed to understand the tradeoffs of this approach.
    \item \OurSys was designed to efficiently serve an index with relatively high traffic. However, further storage tiering to HDDs may allow efficient search on datasets with many mostly-cold tenants, such as per-user indices.
    \item \OurSys is deployed across many machines in a cloud region. Inter-zone latency within a region can be up to $2$ms, and cross-rack bandwidth is oversubscribed. If \OurSys were deployed on a dense cluster of machines with a fully connected network, performance would improve and it might become feasible to store the compressed vectors in a shared memory pool to reduce storage amplification.
\end{itemize}
\section{Acknowledgments}
We would like to thank Dafan Liu, Gena Tertychnyi, and Adelin Miloslavov who gave valuable performance advice about the key-value store, node scoring service, and orchestration service.
\section*{Impact Statement}

This paper presents work whose goal is to advance the field of 
Machine Learning. There are many potential societal consequences 
of our work, none which we feel must be specifically highlighted here.

\bibliography{main}
\bibliographystyle{icml2025}

\end{document}